\documentclass[%
 aps,prb,showpacs,showkeys,
 amsmath,amssymb,floatfix,
 reprint,%
]{revtex4-1}

\usepackage{graphicx}
\usepackage{dcolumn}
\usepackage{bm}
\usepackage{color}

\usepackage{hyperref}
\hypersetup{
   colorlinks,
   linkcolor={blue},
   citecolor={blue},
   urlcolor={blue} 
}

\usepackage{comment}

\begin{document}

\preprint{AIP/123-QED}

\title[
]{Orbitally limited pair-density wave phase of multilayer superconductors}

\author{David M\"{o}ckli
}
\email[E-mail me at: ]{d.mockli@gmail.com}
\affiliation{
The Racah Institute of Physics, The Hebrew University of Jerusalem, Jerusalem 9190401, Israel}
\affiliation{ 
Theoretische Physik, ETH-Z\"{u}rich, 8093 Z\"{u}rich, Switzerland}
\affiliation{ 
Instituto de F\'{\i}sica, Universidade Federal Fluminense, Niter\'{o}i, RJ, 24.210-340, Brazil}
\author{Youichi Yanase} 
\affiliation{Department of Physics, Graduate School of Science, Kyoto University, Kyoto 606-8502, Japan}
\author{Manfred Sigrist}%
\affiliation{ 
Theoretische Physik, ETH-Z\"{u}rich, 8093 Z\"{u}rich, Switzerland
}%

\date{\today}

\begin{abstract}

We investigate the magnetic field dependence of an ideal superconducting vortex lattice in the parity-mixed pair-density wave phase of multilayer superconductors within a circular cell Ginzburg-Landau approach. In multilayer systems, due to local inversion symmetry breaking, a Rashba spin-orbit coupling is induced at the outer layers. This combined with a perpendicular paramagnetic (Pauli) limiting magnetic field stabilizes a staggered layer dependent pair-density wave phase in the superconducting singlet channel. The high-field pair-density wave phase is separated from the low-field BCS phase by a first-order phase transition. The motivating guiding question in this paper is: what is the minimal necessary Maki parameter $\alpha_M$ for the appearance of the pair-density wave phase of a superconducting trilayer system? To address this problem we generalize the circular cell method for the regular flux-line lattice of a type-II superconductor to include paramagnetic depairing effects. Then, we apply the model to the trilayer system, where each of the layers are characterized by Ginzburg-Landau parameter $\kappa_0$, and a Maki parameter $\alpha_M$. We find that when the spin-orbit Rashba interaction compares to the superconducting condensation energy, the orbitally limited pair-density wave phase stabilizes for Maki parameters $\alpha_M> 10$.

%
\end{abstract}

\pacs{74.20.De, 74.20.Rp, 74.25.Dw, 74.25.Op, 74.25.Uv, 74.70.Tx, 74.78.Fk}
\keywords{Ginzburg-Landau theory, multilayer superconductors, magnetization curves, Rashba spin-orbit-coupling, parity-mixed superconductivity, pair-density wave phase, paramagnetic limiting, vortex lattice.}
                          
\maketitle

\section{Introduction}

Magnetic fields applied to superconductors with Cooper pairing in the spin-singlet channel are in two ways detrimental for the superconducting phase. The first is through the coupling to the charge which confines electrons into cyclotron orbits leading to "orbital depairing". The second originates from the Zeeman coupling to the spin by breaking up the spin singlet configuration of the Cooper pair, called "paramagnetic limiting". The corresponding orbital and paramagnetic upper critical fields  are denoted by $ H_{c2}$ and $H_p $, respectively.
In most superconductors the latter is irrelevant because superconductivity disappears at $ H_{c2} $, much smaller than $ H_p $. However, in the opposite limit remarkable features may appear such as the famous Fulde-Ferrell-Larkin-Ovchinnikov (FFLO) state in high magnetic fields if $ H_{c2}$ sufficiently exceeds $ H_p $, quantified by the Maki parameter $\alpha_M = \sqrt{2}H_{c2}(0)/H_p(0)$\cite{Fulde1964,Larkin1964,Maki1966,Matsuda2007}. 
Among the materials where the realization of such an FFLO state is suspected is the heavy fermion superconductor CeCoIn$_5$\cite{Bianchi2003}. 
However, it turned out that  this superconductor is more complex, because the new phase that appears at low temperatures and high magnetic fields
 has also spin magnetic order, the so-called Q-phase \cite{Kenzelmann2008}.

\begin{figure}
\centering
\includegraphics[width=0.45\textwidth]{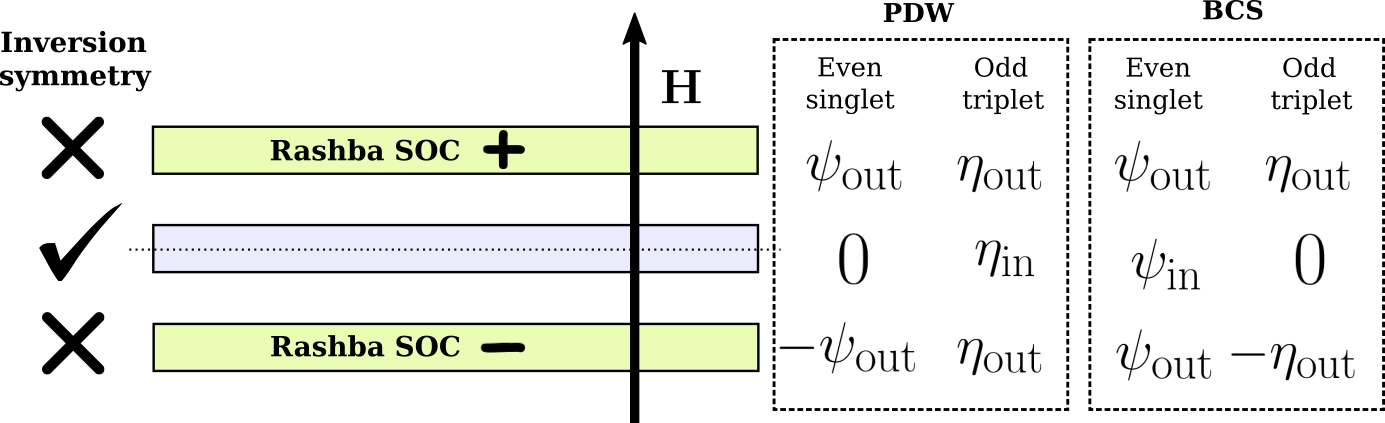}
\caption{\label{mirror} Superconducting trilayer system motivated from artificial Kondo superlattices. Inversion symmetry is locally broken at the outer layers, whereas the inner layer remains inversion symmetric, both globally and locally. Inversion symmetry breaking induces Rashba effects at the outer layers with opposite directions. The PDW and BCS are the two order parameter solutions allowed by symmetry. The PDW state is favorable at high magnetic fields. }
\end{figure}

Recently, Shishido et al produced artificial superlattices consisting of regular stacks of several layers of CeCoIn$_5$ alternating with several layers of YbCoIn$_5$, in
 this way separating the layers of the heavy Fermion material by a normal metal\cite{Shishido2010,Mizukami2011}.
For superlattices where the stacks of CeCoIn$_5$ contain three or more layers, the system remained superconducting. 
It was found that these systems are unusually robust against magnetic fields\cite{Shishido2010,Goh2012,Shimozawa2014,Shimozawa2016}. 
It has been suggested that this feature might be connected with reduced symmetry in the superlattice, so-called local non-centrosymmetricity\cite{Yoshida2012}.
A particularly simple but highly interesting example for this is the trilayer system (see Fig.\ref{mirror}), where the middle layer has inversion symmetry, 
as it constitutes a mirror plane for the system, while the outer layers have a different environment above and below. 
This lack of local mirror symmetry leads to Rashba-type spin-orbit coupling which induces parity-mixing for the Cooper paring states and at the
 same time reduces the effect of paramagnetic limiting for fields perpendicular to the layers\cite{Frigeri2004}.
Soon it was recognized that this structure could give rise to various intriguing properties in a magnetic field, 
such as a complex stripes phase related to the FFLO state, crystalline topological superconductivity and pair density wave phases among other features\cite{Yoshida2012,Maruyama2012,Maruyama2013,Yoshida2013,Yoshida2014,Watanabe2015,Yoshida2015a,Higashi2016}. 
The important ingredient for this is local non-centrosymmetricity and a large Maki parameter.  

Among these exotic phases, we will focus here on the field-induced pair density wave (PDW) phase, characterized below, which is supposed to appear at low temperature for sufficiently high magnetic fields perpendicular to the layers.
This proposal is substantiated by a microscopic model calculation on the level of a Bogolyubov-de Gennes formulation for a trilayer system (see Fig.\ref{mirror}), where the electrons only couple through their spin to the magnetic field, such that only paramagnetic limiting destroys Cooper pairs\cite{Yoshida2012}. In our study we would like to extend the approach to include the mixed phase with the vortex lattice and the effects of orbital depairing. For this purpose we formulate the equivalent Ginzburg-Landau (GL) theory for the trilayer system and analyze it approximating the Abrikosov vortex lattice by generalizing the so-called circular cell method invented and frequently used in the context of vortex matter\cite{Clem1975,Brandt2009,Hao1991,Pogosov2001}. In this way we are able to probe the influence of different depairing mechanisms on the stability of the PDW phase and eventually provide some semiquantitative assessment of the situation in CeCoIn$_5$/YbCoIn$_5$ superlattices.

In this paper we start with the formulation of the GL functional for a monolayer spin-singlet superconductor in order to examine the effect of paramagnetic limiting on the mixed phase, where we treat the vortex lattice with the circular cell method. This system is centrosymmetric such that parity-mixing does not occur. After this test case we turn to the trilayer stack, which requires a more involved multi-component order parameter in both spin-channels for every layer which is necessary to capture both the ordinary BCS and the PDW phase. An important quantity in this context is the Maki parameter $ \alpha_M $ that is needed to exceed a certain threshold to allow for the PDW phase. The derivation of our free energy functional and related discussions will be supported by the three sections in the appendix. 

\section{Ginzburg-Landau theory \label{sec1}}

\subsection{Free energy of a monolayer superconductor}

In this section we develop a Ginzburg-Landau-type model for an ideal vortex lattice of a high-$\kappa$ type-II superconductor, where paramagnetic depairing is included. In order to treat the vortex lattice, we employ the Wigner-Seitz approximation which uses a circular vortex unit cell, neglecting effects due to a specific vortex lattice geometry \cite{Pogosov2001}. This allows us to provide analytical calculations to a large extent.

The PDW phase we would like to address is expected to be realized at high magnetic fields and low temperatures. Therefore, we focus our discussion to the  $T=0$ limit, where a simple relation between the normal state susceptibility $\chi_n$ and the Maki parameter $\alpha_M$ applies (see appendix \ref{apa}).
The zero-temperature free energy density in terms of dimensionless quantities (see appendix \ref{apa} for more details) can be written as\cite{Pogosov2001,P.Mineev2012}

\begin{equation}
\begin{split}
\mathcal{F} =  & \frac{1}{A_\mathrm{cell}}\int_\circ \mathrm{d}^2\rho\left(-f^2(\rho)+\frac{f^4(\rho)}{2}+\frac{[\nabla f(\rho)]^2}{\kappa_0^2}\right)\\
&+\frac{1}{A_\mathrm{cell}}\int_\circ \mathrm{d}^2\rho\left[f^2(\rho)\left(\frac{\nabla\varphi}{\kappa_0}-\boldsymbol{\mathcal{A}}(\rho)\right)^2+\frac{{\mathcal{B}}^2(\rho)}{1+\chi_n}\right]\\
&+\frac{1}{A_\mathrm{cell}}\int_\circ \mathrm{d}^2\rho\,\chi_n\mathcal{B}^2(\rho) f^2(\rho),
\end{split}
\label{freelong}
\end{equation}
or $\mathcal{F}=\mathcal{F}_c+\mathcal{F}_m+\mathcal{F}_p$ for the three terms,
where $\mathcal{F}_c$, $\mathcal{F}_m$ and $\mathcal{F}_p$ refer to the vortex core energy,  the orbital magnetic coupling and the paramagnetic limiting term, respectively. Here $f(\rho)$ is the renormalized superconducting order parameter that due to circular symmetry only depends on the radial coordinate $ \rho $ of the vortex cell, where
$A_\mathrm{cell}=2\pi/(\kappa_0 \bar{\mathcal{B}})$ is the cell area with $\bar{\mathcal{B}}$ as the mean magnetic induction. The magnetic induction can be written in terms of a vector potential $\boldsymbol{\mathcal{B}} = \nabla\times\boldsymbol{\mathcal{A}}$, and $\kappa_0$ is the standard Ginzburg-Landau parameter ignoring correction due to the paramagnetic depairing. The normal state susceptibility
$\chi_n$ determines the strength of the paramagnetic effect via $\chi_n = (\alpha_M/\kappa_0)^2/2$ involving the Maki parameter $ \alpha_M $ (see equation \eqref{maki}).
In dimensionless units the flux quantum threading a unit cell reads $\phi_0 = 2\pi/\kappa_0$, and the vortex unit cell radius is $\rho_\mathcal{B}^2=2/(\kappa_0\bar{\mathcal{B}})$.

The dimensionless magnetic field $\mathcal{H}$ and magnetic induction $\mathcal{B}$ are related by
\begin{equation}
\mathcal{H}=\frac{1}{2}\frac{\partial \mathcal{F}}{\partial \bar{\mathcal{B}}},
\label{hfield}
\end{equation}
from which we extract the magnetization $\mathcal{M}=\mathcal{B}-\mathcal{H}$ and magnetic susceptibility $\chi=\mathcal{M}/\mathcal{H}$.

\subsection{Variational circular cell procedure}

The usual Ginzburg-Landau scheme follows a variational minimization of the free energy functional \eqref{freelong} with respect to $f(\rho)$ and $\boldsymbol{\mathcal{A}}$. Here, for the sake of analytical insights and simplicity, we employ a procedure proposed by Clem \cite{Clem1975}, which uses the variational Ansatz
\begin{equation}
f^2(\rho)=f^2_\infty \frac{\rho^2}{\rho^2+\xi_c^2}
\label{ansatz}
\end{equation}
to model the vortex core,
where $f_\infty$ and $\xi_c$ are variational parameters for the bulk magnitude of the order parameter and vortex core size, respectively. This  Ansatz eliminates one of the two Ginzburg-Landau equations and provides one remaining differential equation that can be solved analytically. Substituting the Ansatz \eqref{ansatz} into the free energy \eqref{freelong} the variational derivative with respect to $\boldsymbol{\mathcal{A}}$ leads to 

\begin{equation}
\frac{\nabla^2\boldsymbol{\mathcal{A}}}{1+\chi_n}+f_p^2\frac{\rho^2}{\rho^2+\xi_p^2} \left(\frac{\nabla\varphi}{\kappa_0}-\boldsymbol{\mathcal{A}}\right)=0 ,
\label{avec}
\end{equation}
where
\begin{equation}
f_p^2=\frac{f_\infty^2}{1+\chi_n f_\infty^2},\quad\mbox{and}\quad \xi_p^2=\frac{\xi_c^2}{1+\chi_n f_\infty^2}.
\label{noobs}
\end{equation}

After the gauge transformation $\boldsymbol{\mathcal{A}}\rightarrow \boldsymbol{\mathcal{A}} +\nabla\varphi/\kappa_0$, we take the curl of equation \eqref{avec}, and use it to eliminate $\boldsymbol{\mathcal{A}}$ to obtain \cite{saint1969type}
\begin{equation}
\frac{1}{1+\chi_n}\frac{1}{\rho}\frac{\mathrm{d}}{\mathrm{d}\rho}\left[\frac{1}{f_p^2}\frac{\rho^2+\xi_p^2}{\rho}\frac{\mathrm{d}\mathcal{B}}{\mathrm{d}\rho}\right]=\mathcal{B}.
\label{diffeq}
\end{equation}
This is a modified Bessel differential equation whose solution determines the spatial distribution of the magnetic induction $\mathcal{B}$.
A general solution of equation \eqref{diffeq} is \cite{Pogosov2001}
\begin{equation}
(1+\chi_n)\mathcal{B}(\rho)=c_1 K_0\left(f_p \sqrt{\rho^2+\xi_p^2}\right)+c_2 I_0\left(f_p \sqrt{\rho^2+\xi_p^2}\right),
\label{induction}
\end{equation}
where $I_n(x)$ and $K_n(x)$ are $n^{\mathrm{th}}$-order modified Bessel functions of the first and second kind, respectively. The coefficients $c_1$ and $c_2$ are determined by the conditions that the flux threading a unit cell (for the $\alpha_M=\chi_n=0$ case) is a flux quantum and that the current density is zero at the cell boundary. Defining $\mathrm{P}^2=2/(\kappa_0\bar{\mathcal{B}})+\xi_p^2$, these coefficients can be expressed analytically as
\begin{equation}
c_1=\frac{f_\infty}{\kappa_0\xi_c}\frac{I_1(f_p\mathrm{P})}{K_1(f_p\xi_p )I_1(f_p\mathrm{P})-I_1(f_p\xi_p)K_1(f_p\mathrm{P})},
\end{equation}
\begin{equation}
c_2=\frac{f_\infty}{\kappa_0\xi_c}\frac{K_1(f_p\mathrm{P})}{K_1(f_p\xi_p )I_1(f_p\mathrm{P})-I_1(f_p\xi_p)K_1(f_p\mathrm{P})}.
\end{equation}

Inserting $ f(\rho) $ and $ \mathcal{B}(\rho) $ in Eq.\eqref{freelong} we perform the integrals over $ \rho $ to obtain
\begin{equation}
\begin{split}
\mathcal{F}_c = &+ \frac{1}{2}\left(1-f_\infty^2\right)^2+\frac{f_\infty^4}{2} \\
& +\frac{1}{2}\bar{\mathcal{B}}\kappa_0\xi_c^2 f_\infty^2(1-f_\infty^2)\ln\left(1+\frac{2}{\bar{\mathcal{B}}\kappa_0\xi_c^2}\right) \\
& +\frac{\bar{\mathcal{B}} f_\infty^2(1+\bar{\mathcal{B}}\kappa_0\xi_c^2)}{\kappa_0(2+\bar{\mathcal{B}}\kappa_0\xi_c^2)^2} -\frac{f_\infty^4}{2+\bar{\mathcal{B}}\kappa_0\xi_c^2}.
\end{split}
\label{intcore}
\end{equation}
Note that $\bar{\mathcal{B}}\kappa_0\xi_c^2/2=(\xi_c/\rho_\mathcal{B})^2$.
To calculate $\mathcal{F}_m$ we use the result by Hao-Clem \cite{Hao1991} that states $\mathcal{F}_m=\bar{\mathcal{B}}\mathcal{B}(0)$ leading to
\begin{equation}
\mathcal{F}_m=\frac{\bar{\mathcal{B}}}{1+\chi_n}\frac{f_\infty}{\kappa_0\xi_c}\frac{K_0(f_p\xi_p)I_1(f_p\mathrm{P})+I_0(f_p\xi_p)K_1(f_p\mathrm{P})}{K_1(f_p\xi_p )I_1(f_p\mathrm{P})-I_1(f_p\xi_p)K_1(f_p\mathrm{P})}.
\label{eq:hcapp}
\end{equation}
If the order parameter is very small, $f_\infty\ll 1$, one can approximate eq. \eqref{eq:hcapp} simply by the lowest order term, $\mathcal{F}_m^{(0)}=\bar{\mathcal{B}}^2/(1+\chi_n)$. 

To evaluate the last term $\mathcal{F}_p$ we use the mean value of the magnetic induction $\bar{\mathcal{B}}$  
\begin{equation}
\mathcal{F}_p=\bar{\mathcal{B}}^2 f_\infty^2\mathcal{Q}(0)\left[1+\frac{\bar{\mathcal{B}}\kappa_0 \xi_c^2}{2}\ln\left(1-\frac{2}{2+\bar{\mathcal{B}}\kappa_0\xi_c^2}\right)\right],
\label{intpar}
\end{equation}
where $ \mathcal{Q}(T) $ is defined in Appendix \ref{apa} and $ \mathcal{Q}(0) = \chi_n/(1+\chi_n) $.

The resulting free energy summing up all three terms, $\mathcal{F}[f_\infty,\xi_c]$,  is an analytic free energy density for a vortex lattice subject to paramagnetic limiting. The minima with respect to the variational parameters $(f_\infty,\xi_c)$ for some values of $\bar{\mathcal{B}}$ can be obtained by simultaneous numerical minimization of the free energy density. Here we use a Nelder-Mead procedure to find $(f_\infty,\xi_c)_{\rm min}$. We include the expressions for the magnetic field $\mathcal{H}$ and the lower critical field $\mathcal{H}_{c1}$ in Appendix \ref{abig}.

\subsection{Results of a numerical evaluation}

\begin{figure}
\includegraphics[width=0.5\textwidth]{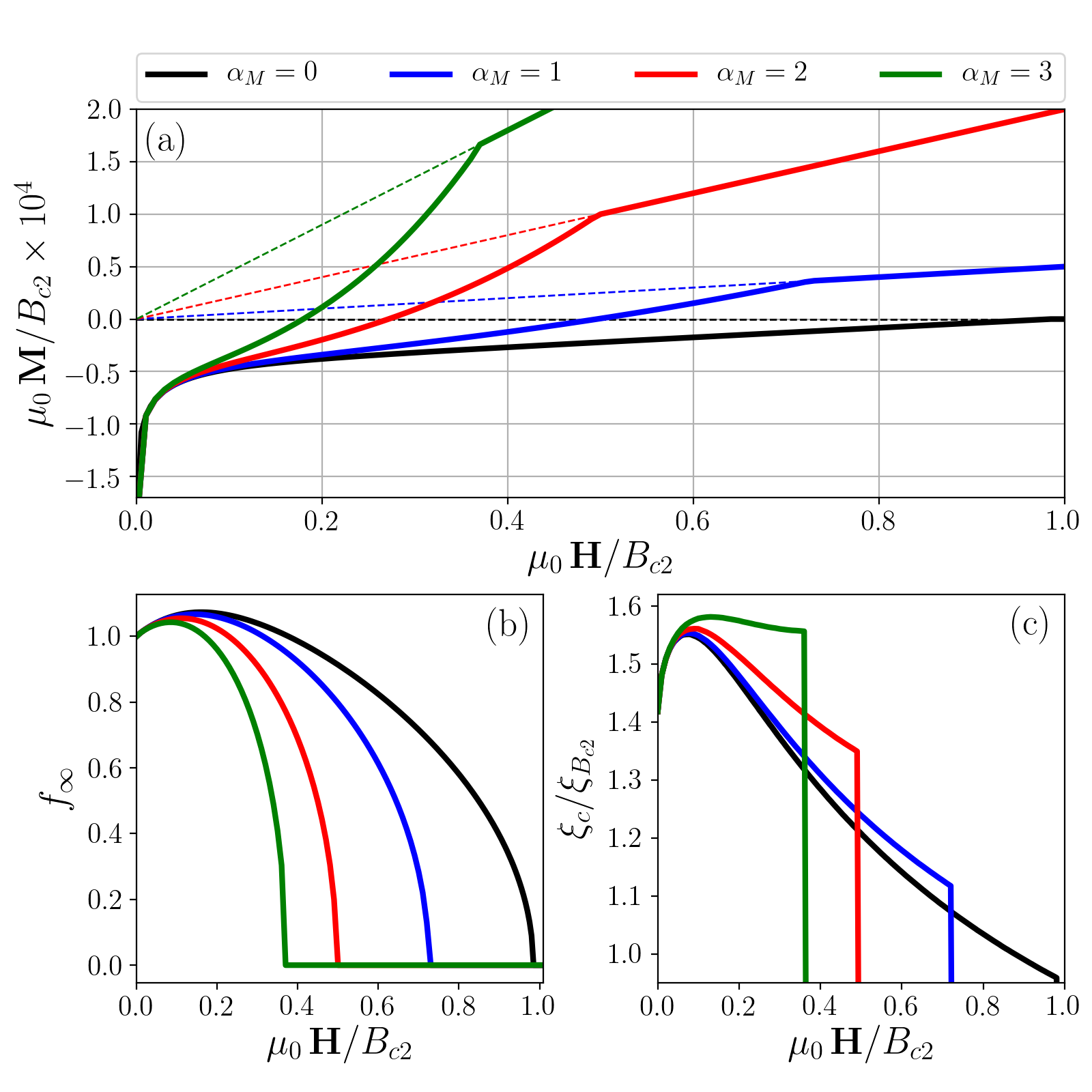}
\caption{\label{fig:plot1}(a) Magnetization curves at $T=0$ staring from a $\kappa_0=100$ superconductor for different Maki parameters. The magnetic field and the magnetization are given in units of the orbital upper critical field $B_{c2} = \kappa_0 $. (b) Dependence of $f_\infty$ on the $H$-field for different Maki parameters.  (c) Dependence of the variational parameter $\xi_c$ (modeling the vortex core size) on the $H$-field for different Maki parameters. $\xi_{B_{c2}}$ is the upper critical core size for $\alpha_M=0$.}
\end{figure}

We consider now the behavior of the mixed phase for different values of the Maki parameter, $\alpha_M = \{0,1,2,3\}$, providing the results for the magnetization, the order parameter and the vortex core size as a function of the applied magnetic field in Fig.\ref{fig:plot1}. The Ginzburg-Landau parameter is $ \kappa_0 = 100 $ such that the lower critical field is very small and will be ignored here. Fig.\ref{fig:plot1}(a) shows the magnetization curves where the upper critical field, defined by the vanishing of $ f_{\infty} $, shrinks with increasing $ \alpha_M $. The normal state magnetization extrapolates to zero and is enhanced for increasing $ \alpha_M $ together with the spin contribution through $ \chi_n $, as explained in Eq.\eqref{maki} of Appendix \ref{apb}. The qualitative features of the magnetization curves agree with the experimental results observed in CeCoIn$_5$\cite{Tayama2002,Tayama2002b} or  KFe$_2$As$_2$ \cite{Burger2013}.
A similar behavior can also be derived from a microscopic approach (see for example Ref.\onlinecite{Ichioka2011}). Note that the rather singular response of $ M $ in the limit of $ H = 0 $ appears because we ignore the lower critical field which would truncate this behavior.

Fig. \ref{fig:plot1}(b) displays the field dependence of the order parameter $ f_{\infty} $. The feature of the initial rise of $ f_{\infty} $ exceeding $1$ is an artifact of our circular cell approach \cite{Pogosov2001}. The same spurious increase is also observed in Fig.\ref{fig:plot1}(c) for the vortex core radius  $\xi_c $ which we renormalize with $ \xi_{B_{c2}} = 1/ \kappa_0 $, the core radius at $ B_{c2} $ for $ \alpha_M=0 $. Thus, the more reliable behavior is found rather for higher magnetic fields where $ f_{\infty} $ is gradually suppressed. It is interesting to note that the core radius increases for finite $ \alpha_M $ relative to $ \alpha_M =0 $ indicating that the paramagnetic depairing acts detrimental to the order parameter within the vortex core, where the magnetic field is also largest and the order parameter weakest.  

\section{Pair-density wave phase in the trilayer system \label{sec2}}

We now turn to the trilayer system as introduced at the beginning. The properties of this system under various conditions have been investigated using a Bogolyubov-de Gennes (BdG) formulation restricting to paramagnetic depairing \cite{Yoshida2012,Yoshida2015a,Yoshida2014,Maruyama2012}. In this section we introduce the GL formulation in order to include orbital depairing as we did for the monolayer case above. 

\subsection{Local inversion symmetry breaking}

In a system with time reversal symmetry and inversion symmetry, we categorize superconducting phases by the Cooper pair symmetry into even-parity spin-singlet and odd-parity spin-triplet pairing states\cite{Sigrist1991}. In non-centrosymmetric systems the lack of inversion deprives Cooper pairs of having a definite parity, which, hence, can be considered as parity-mixed states\cite{Hayashi2006,Yip2014}. 

A superconducting system composed of a stack of three identical layers can be regarded as a locally non-centrosymmetric system (see Fig.\ref{mirror}). 
Globally the trilayer system is centrosymmetric as there are inversion centers in the middle layer which is also a mirror plane. This middle layer constitutes, therefore, also a centrosymmetric subsystem where the standard classification of order parameter symmetries applies. This is different for the outer-layers that do have different environments above and below which we consider as non-centrosymmetric subsystems whose superconducting order parameters are subject to parity mixing. The superconducting order parameters of the outer layers are related by symmetry. 

Parity-mixing is induced by spin-orbit coupling that is of Rashba-type for these outer layers with opposite sign for the upper and lower one (see Fig.\ref{mirror}). The dominant order parameter is in our case even-parity spin-singlet in character and for the three layers shall be represented by $ \psi_i = \{\psi_{\mathrm{out}},\psi_{\mathrm{in}},\psi_{\mathrm{out}}\}$ ($ i=1,2,3$ where 1 and 3 denote the outer and 2 the middle layer). The admixed odd-parity component only appears in the outer layers with opposite sign, $\eta_i = \{\eta_{\mathrm{out}},0,-\eta_{\mathrm{out}}\}$. It is important to note that this spin-triplet component has a spin configuration corresponding to equal-spin pairing with spin orientation perpendicular to the plane, making it robust against spin polarization along this direction. This combination of these two order parameters we will call in the following BCS-phase following the nomenclature of Ref.\onlinecite{Yoshida2012}. In this reference it was shown that in high magnetic fields (perpendicular to the layers) a phase dubbed pair density wave (PDW) is stabilized with the order parameter configuration $\psi_i = \{\psi_{\mathrm{out}},0,-\psi_{\mathrm{out}}\}$ and $\eta_i = \{\eta_{\mathrm{out}},\eta_{\mathrm{in}},\eta_{\mathrm{out}}\}$, where the spin-singlet component changes sign between the two outer layers. In the following we will formulate a GL theory which incorporates the ingredients to reproduce this phenomenology.

\subsection{Free energy of the trilayer system}

We now extend the GL theory developed in the preceding section \ref{sec1} for a single superconducting layer to the case of three layers in order to describe the occurrence of a PDW state in a magnetic field and to estimate the necessary Maki parameter $\alpha_M$. In order to integrate the parity-mixing effect we need six order parameters, three spin-singlet components, $\Psi_i(r,\alpha_i)=\psi_i(r)\exp i\alpha_i$, and three triplet components, $\Xi_i(r,\beta_i)=\eta_i(r)\exp i\beta_i$.
Again, we write the free energy in terms of dimensionless quantities that will be indicated by the argument $\psi_i(r)\rightarrow \psi_i(\rho)$ within the single-vortex cell, whereby we renormalize in the same way as above and explained in Appendix \ref{apa}. We write the dimensionless zero temperature free energy density as the sum of five terms, 
\begin{equation}
\mathcal{F} = \mathcal{F}_c + \mathcal{F}_\epsilon +\mathcal{F}_J+\mathcal{F}_m+\mathcal{F}_p,
\end{equation}
which refer respectively to the core, order parameter mixing, Josephson inter-layer coupling, magnetic orbital part, and paramagnetic term. Each one of the terms will be introduced and discussed now.

The core free energy density including all order para\-meter components is given by
\begin{equation}
\begin{split}
\mathcal{F}_c = \frac{1}{A_{\mathrm{cell}}}\int_\circ \mathrm{d}^2\rho \sum_{i}\biggr(  & -\psi_i^2(\rho)+\frac{\psi_i^4(\rho)}{2}+\frac{\left[\nabla\psi_i(\rho)\right]^2}{\kappa_0^2} \\
& +\frac{\eta_i^4(\rho)}{2}+\frac{\left[\nabla\eta_i(\rho)\right]^2}{\kappa_0^2}\biggr),
\end{split}
\label{fc3}
\end{equation}
where $A_{\mathrm{cell}}=\pi\rho_\mathcal{B}^2$ is the vortex cross section unit cell area. We assume that the threading vortex line is vertically aligned throughout the three superconducting layers. Both $\psi_i(\rho)$ and $\eta_i(\rho)$ are the dimensionless superconducting order parameters, renormalized with respect to the maximum layer values. In order to be able to perform to a large extent analytical calculations, we impose several simplifications here.
The fourth-order mixed terms are neglected. 
The spin-triplet square term $\eta_i^2(\rho)$ has been omitted for simplicity fixing the bare critical temperature of this pairing channel to zero in all layers. A finite $ T_c $ for $ \eta_i $ would not alter our conclusions qualitatively. We also fixed the GL parameter $\kappa_0$ to be the same for both the spin-singlet and -triplet components which implies that the orbital depairing is equally detrimental to both. 

The parity mixing induced by the  Rashba spin-orbit coupling in layers $1$ and $3$ is implemented by a second-order mixing term 
\begin{equation}
\mathcal{F}_\epsilon=\frac{1}{A_{\mathrm{cell}}}\int_\circ \mathrm{d}^2\rho \sum_{i}\epsilon_i\gamma\,\psi_i(\rho)\eta_i(\rho)\,\cos(\alpha_i-\beta_i).
\end{equation}
with $\epsilon_i=(1,0,-1)$ reflecting that the spin-orbit coupling has opposite sign on the two sides in accordance with the mirror symmetry with respect to the middle layer. The parameter $\gamma$ is the coupling strength. In previous studies of the trilayer system, neglecting the orbital depairing effect, phase differences $\alpha_i-\beta_i$ were constant and either $0$ or $\pi$ \cite{Maruyama2012,Yoshida2012,Yoshida2014}. Here we assume that we can keep this registry of the phases even including the mixed phase through sufficiently strong interlayer Josephson coupling,
\begin{equation}
\begin{split}
\mathcal{F}_J=\frac{J}{A_{\mathrm{cell}}}\int_\circ \mathrm{d}^2\rho \sum_{\langle i,j\rangle}\bigr( & |\psi_i(\rho)-\psi_j(\rho)|^2\\
&+|\eta_i(\rho)-\eta_j(\rho)|^2\bigr),
\end{split}
\end{equation}
where $ J $ is the coupling strength and $\langle i,j\rangle$ indicates the summation over neighboring layers, i.e. $ (1,2) $ and $ (2,3) $. 
\textcolor{black}{In the following we assume the coupling constants $ J $ and $ \gamma $ to be comparable to the condensation energy density, $\mathcal{F}_{\mathrm{cond}} = -|\psi|^2+|\psi|^4/2$ which is $ -1/2 $ for the minimizing order parameter $ | \psi | = 1 $.} 

Next we consider the magnetic orbital part. A dramatic simplification occurs, if we allow the magnetic induction $\mathcal{B}_i(\rho)$ to be identical in all layers, $\mathcal{B}(\rho)$. Together with the constant phase differences this allows also to perform the analog gauge transformation for the vector potential as done for the monolayer case above (Sec.\ref{sec1}) because $ \alpha_i - \beta_i $ implies $\nabla\alpha_i=\nabla\beta_i$. Then, the orbital magnetic term can be written as
\begin{equation}
\mathcal{F}_m = \frac{1}{A_{\mathrm{cell}}}\int_\circ \mathrm{d}^2\rho \sum_{i}\left(\left[\psi_i^2(\rho)+\eta_i^2(\rho)\right]\mathcal{A}^2+\frac{\mathcal{B}^2(\rho)}{1+\chi_n}\right),
\end{equation}
where $ \chi_n $ is the susceptibility parameter, identical for all layers. 

At last, we write the paramagnetic coupling to the superconducting order parameter as,
\begin{equation}
\mathcal{F}_p = \frac{1}{A_{\mathrm{cell}}}\int_\circ \mathrm{d}^2\rho \,\chi_n\,\bar{\mathcal{B}}^2 \sum_i d_i \psi_i^2(\rho),
\end{equation}
which only affects the spin-singlet component. Here we introduce the layer-dependent renormalization factor $ d_i $ which is $ d_2 = 1 $ but $ d_{1,3} \ll 1 $. The two outer layers have a strongly reduced renormalization factor $ d_i $ because Rashba spin-orbit coupling turns the Pauli spin polarization into a van Vleck type spin polarization which does not lead to pair breaking\cite{Maruyama2012}.

\subsection{Variational circular cell procedure}

Following the previous variational treatment, we take the Ansatz for the order parameters
\begin{equation}
\psi_i^2(\rho)=\psi_{i \infty}^2 \frac{\rho^2}{\rho^2+\xi_c^2} ; \quad \eta_i^2(\rho)=\eta_{i \infty}^2 \frac{\rho^2}{\rho^2+\xi_c^2}
\end{equation}
and minimize with respect to $\boldsymbol{\mathcal{A}}$ obtaining an analogous equation to \eqref{avec}
\begin{equation}
\frac{\nabla^2\boldsymbol{\mathcal{A}}}{1+\chi_n}=\Delta_\infty^2 \,\frac{\rho^2}{\rho^2+\xi^2_\Delta}\boldsymbol{\mathcal{A}},
\end{equation}
where
\begin{equation}
\Delta_\infty^2=\frac{\sum_i(\psi_{i \infty}^2+\eta_{i \infty}^2)}{1+\chi_n\psi_{2\infty}^2};\quad \xi^2_\Delta = \frac{\xi_c^2}{1+\chi_n\psi_{2\infty}^2}
\end{equation}
is the representative variational order parameter and vortex core size of the trilayer system. Defining the quantity
\begin{equation}
C_{\kappa}(\bar{\mathcal{B}},\xi_c)=1+\frac{\bar{\mathcal{B}}\kappa_0\xi_c^2}{2}\ln\left(1-\frac{2}{2+\bar{\mathcal{B}}\kappa_0\xi_c^2}\right),
\end{equation}
for the sake of compactness, 
the integrated free energy density for the vortex cell is then written as
\begin{equation}
\mathcal{F}[\psi_{i \infty},\eta_{i \infty},\xi_c] = \mathcal{F}_c+\mathcal{F}_\epsilon+\mathcal{F}_J+\mathcal{F}_m+\mathcal{F}_p,
\label{full3l}
\end{equation}
with
\begin{equation}
\begin{split}
\mathcal{F}_c = & -C_{\kappa}(\bar{\mathcal{B}},\xi_c)\sum_i\psi_{i \infty}^2\\
&+\left[C_{\kappa}(\bar{\mathcal{B}},\xi_c)-\frac{1}{2+\bar{\mathcal{B}}\kappa_0\xi_c^2}\right]\sum_i\left(\psi_{i \infty}^4+\eta_{i \infty}^4\right)\\
&+\frac{\bar{\mathcal{B}}(1+\bar{\mathcal{B}}\kappa_0\xi_c^2)}{\kappa_0(2+\bar{\mathcal{B}}\kappa_0\xi_c^2)^2}\sum_i\left(\psi_{i \infty}^2+\eta_{i \infty}^2\right),
\end{split}
\end{equation}
\begin{equation}
\mathcal{F}_\epsilon= C_{\kappa}(\bar{\mathcal{B}},\xi_c)\sum_i \epsilon_i\,\gamma\psi_{i \infty}\eta_{i \infty}\,\mathrm{sgn}_i(0,\pi),
\end{equation}
\begin{equation}
\mathcal{F}_J = C_{\kappa}(\bar{\mathcal{B}},\xi_c) \sum_{\langle i,j\rangle}J\left(|\psi_{i \infty}-\psi_{j \infty}|^2|+|\eta_{i \infty}-\eta_{j \infty}|^2\right),
\end{equation}
\begin{equation}
\begin{split}
\mathcal{F}_m & =  \frac{\bar{\mathcal{B}}}{1+\chi_n}\frac{\sqrt{\sum_i(\psi_{i \infty}^2+\eta_{i \infty}^2)}}{\kappa_0\xi_c}\times\\
& \times \frac{K_0(\Delta_\infty\xi_\Delta)I_1(\Delta_\infty\mathrm{P})+I_0(\Delta_\infty\xi_\Delta)K_1(\Delta_\infty\mathrm{P})}{K_1(\Delta_\infty\xi_\Delta )I_1(\Delta_\infty\mathrm{P})-I_1(\Delta_\infty\xi_\Delta)K_1(\Delta_\infty\mathrm{P})},
\end{split}
\end{equation}
\begin{equation}
\mathcal{F}_p = \chi_n\bar{\mathcal{B}}^2 \sum_i d_i \psi_{i \infty}^2,
\end{equation}
and $\mathrm{P}^2=2/(\kappa_0\bar{\mathcal{B}})+\xi_\Delta^2$. 

The trilayer free energy density functional $\mathcal{F}[\psi_{i \infty},\eta_{i \infty},\xi_c]$ has seven variational parameters with respect to which it has to be simultaneously minimized. We may, however, impose the structure of the states allowed by symmetry, that is, the low-field BCS state as
\begin{equation}
\psi_{i \infty}^{\mathrm{BCS}}=\{\psi_{\mathrm{out}},\psi_{\mathrm{in}},\psi_{\mathrm{out}}\},\quad\eta_{i \infty}^{\mathrm{BCS}} = \{\eta_{\mathrm{out}},0,-\eta_{\mathrm{out}}\},
\end{equation}
and the high-field PDW state as
\begin{equation}
\psi_{i \infty}^{\mathrm{PDW}}=\{\psi_{\mathrm{out}},0,-\psi_{\mathrm{out}}\},\quad\eta_{i \infty}^{\mathrm{PDW}} = \{\eta_{\mathrm{out}},\eta_{\mathrm{in}},\eta_{\mathrm{out}}\},
\label{PDWstructure}
\end{equation}
which simplifies the problem considerably. Thus, the free energy functionals can be written for the two states using the corresponding free variational parameters:  $\mathcal{F}_{\mathrm{BCS}}[\psi_{\mathrm{out}},\psi_{\mathrm{in}},\eta_{\mathrm{out}},\xi_c]$ and $\mathcal{F}_{\mathrm{PDW}}[\psi_{\mathrm{out}},\eta_{\mathrm{out}},\eta_{\mathrm{in}},\xi_c]$. Now again we resort to numerical minimization and compare the free energy densities to decide which of the two states is more stable under given conditions, i.e. the mean magnetic induction $\bar{\mathcal{B}}$. Our setup is chosen so that all variational parameters remain real in this procedure. 

Note that by imposing the PDW order parameter structure \eqref{PDWstructure} into the free energy \eqref{full3l}, one immediately sees that the PDW phase is predominantly limited by orbital depairing, whereas the BCS is dominated by paramagnetic limiting, as we will see in the following. 

\begin{figure}
\includegraphics[width=0.5\textwidth]{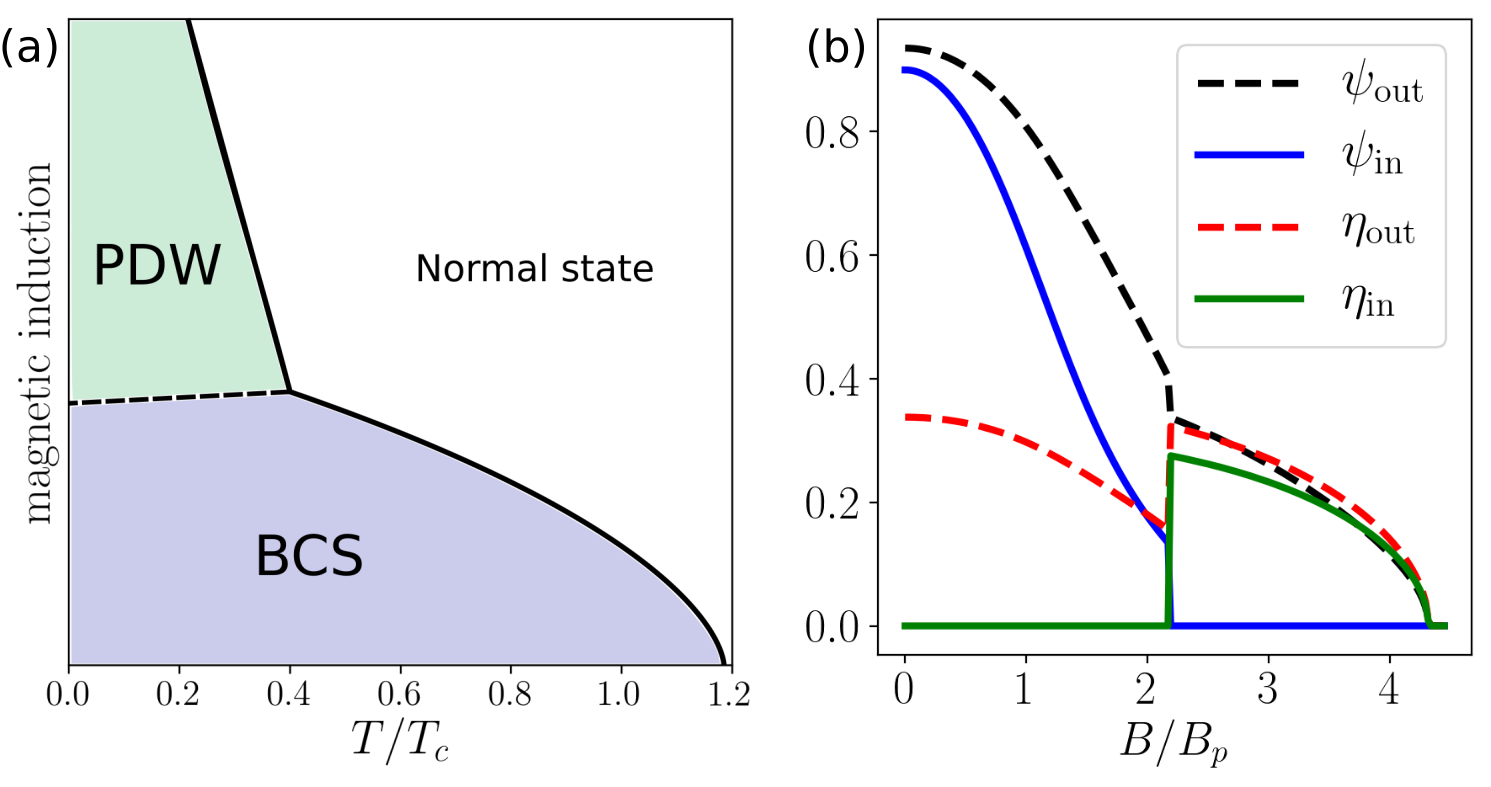}
\caption{\label{fig:pdw} (a) Typical temperature -- magnetic induction phase diagram for a trilayer system with infinite Maki parameter obtained within Ginzburg-Landau theory. Here $T_c$ denotes the critical temperature of the spin-singlet monolayer superconductor. $T/T_c$ closes at $1.2$, which shows a slight increase of the effective critical temperature due to the order parameter mixing even at $B=0$. (b) Cut at $T/T_c=0.27$ showing the magnetic induction dependence of the order parameters. $B_p$ is the monolayer Pauli critical field.}
\end{figure}

\section{Results and Discussion}

For the sake of clarity we extract the role of the different depairing effects by analyzing different limiting cases in sequence. 
First, we discuss the emergence of the PDW phase in the trilayer system purely based on paramagnetic limiting (neglecting orbital depairing), which corresponds to the $\alpha_M\rightarrow\infty$ limit. As mentioned above this case was already investigated by previous works within a BdG approach to a microscopic model\cite{Maruyama2012,Yoshida2012,Maruyama2013,Yoshida2013,Yoshida2014,Watanabe2015,Yoshida2015a}. We show that the features obtained previously can easily be observed also within a Ginzburg-Landau approach on a qualitative level. Second, we discuss the $\alpha_M = 0$ case, which corresponds to pure orbital depairing (neglecting the paramagnetic limiting). Although there is no PDW without paramagnetic limiting, this is an instructive example to see crucial differences of the trilayer system compared to the monolayer case. Even though there is no paramagnetic effect with $\alpha_M = 0$, the outer layers admit an admixed triplet order parameter that is insensitive to the paramagnetic effect to the dominant singlet component. The presence of a triplet component introduces some important differences in the behavior of the overall susceptibility of the superconductor. 
Third, we analyze the PDW phase taking both depairing mechanism simultaneously into account, whereby the relative relevance of the paramagnetic effect with respect to the orbital effect can be tuned by the Maki parameter $\alpha_M = \sqrt{2} H_{c2}/H_p$ ($H_{c2}$ is the purely orbital upper critical field, and $H_p$ is the paramagnetic critical field). In Appendix \ref{apb} we show that the Maki parameter can be estimated within the GL approach via $\alpha_M = \kappa_0\sqrt{2\chi_n}$. From this concise formula for $ \alpha_M $ we conclude that strongly type-II superconductors, large $ \kappa_0 $, as well as enhanced susceptibility $ \chi_n $ are most favorable for the occurrence of the PDW phase.

\subsection{Trilayers subjected to paramagnetic limiting}

We first consider the simplest case with paramagnetic limiting only. Since orbital depairing is neglected we can do without the gradient terms in the free energy density and the spatially modulated order parameter,
\begin{equation}
\begin{split}
\mathcal{F} = & \sum_i\left[-\psi_i^2+\frac{\psi_i^4}{2}+\frac{\eta_i^4}{2}+\epsilon_i\gamma\psi_i\eta_i\right]+\chi_n \mathcal{B}^2\sum_i d_i\psi_i^2\\
&+J\sum_{\langle i,j\rangle}\left[(\psi_i-\psi_j)^2+(\eta_i-\eta_j)^2\right]+\frac{\mathcal{B}^2}{1+\chi_n}.
\end{split}
\end{equation}

In order to understand the basic mechanisms at work we consider first the case with $ d_{1,3} = 0 $ such that only the middle layer is subject to paramagnetic limiting.
For a BCS solution with $\psi_i = \{\psi_{\mathrm{out}},\psi_{\mathrm{in}},\psi_{\mathrm{out}}\}$, $\eta_i = \{\eta_{\mathrm{out}},0,-\eta_{\mathrm{out}}\}$, only $\psi_{\mathrm{in}}$ couples to the $\mathcal{B}$-field. Nevertheless, through the interlayer coupling all order parameters decrease when the field is increased, although weaker for the outer layer than the middle layer. On the other hand, the PDW solution with $\psi_i =\{\psi_{\mathrm{out}},0,-\psi_{\mathrm{out}}\}$, $\eta_i = \{\eta_{\mathrm{out}},\eta_{\mathrm{in}},\eta_{\mathrm{out}}\}$ does not have any order parameter coupling to the $\mathcal{B}$-field, because $\psi_2=0$. For this reason the PDW solution is the favored state for high magnetic fields if $\chi_n$ is sufficiently large. 

Now we turn to a more general situation, where also the outer layers suffer paramagnetic limiting, by setting $ d_1 = d_3 =0.01 $. \textcolor{black}{Moreover, to be concrete, we choose $ \chi_n = 0.001 $ and  $\gamma=J=1$.} In Fig. \ref{fig:pdw}(a) we show the phase diagram temperature versus magnetic induction for these parameters in the trilayer system. Note that the onset of superconductivity is higher than the bare $ T_c $ of the spin-singlet component in each layer due to the support by the parity-mixing. 
The high-magnetic field PDW phase is separated from the BCS phase by a first order phase boundary. This first order transition is also clearly visible in Fig.\ref{fig:pdw}(b) for $ T = 0.27 T_c $ depicted. Here all order parameter components show a discontinuity at $B/B_p\approx 2.2$. In the low-field BCS phase 
$ \psi_{\mathrm{in}} $ decreases faster than $ \psi_{\mathrm{out}} $ as expect from the dominance of paramagnetic limiting in the middle layer. In the PDW phase the spin-triplet component takes a leading role and also appears in the middle layer ($ \eta_{\mathrm{in}} $ in exchange with $ \psi_{\mathrm{in}} $ which has completely disappeared). All  order parameters remaining at these high fields turn to zero at the same magnetic field. Note that $ \psi_{\mathrm{in}} $ survives to higher fields than the nominal paramagnetic limiting field $ B_ p $ because of the support of the other layers through the interlayer coupling. 
To obtain the temperature dependencies, we have reversed the change to dimensionless units, as explained in Appendix \ref{apa}.

\begin{figure}
\includegraphics[width=0.5\textwidth]{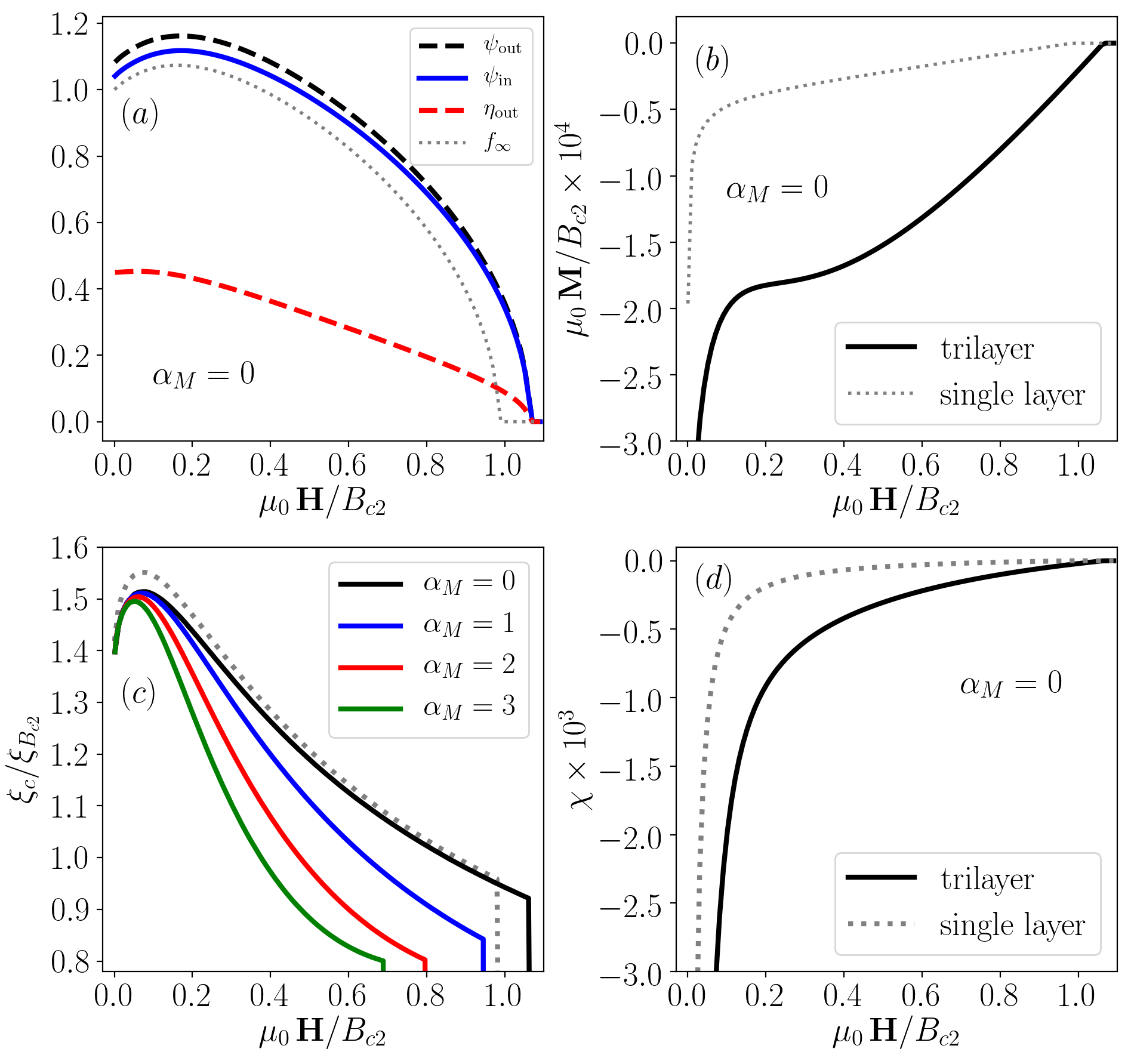}
\caption{\label{fig:comparison} (a) Comparison of a monolayer $\kappa_0=100$ and a trilayer system subjected to pure orbital limiting ($\alpha_M = 0$). The dotted gray curves show the monolayer case (in all panels). In the trilayer system, the order parameter mixing causes a slight increase in the upper critical field. Here we have assumed that both singlet and triplet components are equally orbitally limited, which then yields an additional magnetization due to the triplet components in the trilayer system, which is seen in (b). (c) The suppression of the vortex core size for increasing Maki parameter. This contrasts with the monolayer case, where the core sizes increase. (d) The trilayer system is less susceptible than a monolayer due to the presence of triplet Cooper pairs.}
\end{figure}

\subsection{Trilayers subjected to orbital limiting}

We now address the opposite limit of a trilayer system with orbital depairing only ($\alpha_M=0$), and compare it to the monolayer case of section \ref{sec1}. Because there is no paramagnetic limiting in the middle layer, a PDW phase is out of competition.  Still, the outer layers in the trilayer system will have an induced spin-triplet order parameter, whereas in the monolayer system there is only one singlet component $f_\infty$. In Fig.\ref{fig:comparison} we compare the results for the monolayer to the corresponding trilayer system composed of three such identical layers, where we choose again $\kappa_0=100$. The dotted gray curves show the results already obtained in Fig.\ref{fig:plot1} for the monolayer. In Fig.\ref{fig:comparison}(a) we show the magnetic field dependence of the spin singlet order parameter components together with the induced triplet component $\eta_\mathrm{out}$ at the outer layers due to the order parameter mixing $\psi_\mathrm{out}\eta_\mathrm{out}$ in the outer layers. Again the singlet order parameters and the effective upper critical field are slightly increased with respect to the monolayer. A more pronounced distinction occurs in the magnetization curve in Fig.\ref{fig:comparison}(b). The trilayer system has now parity-mixed superconducting vortices, with both singlet and triplet components equally orbitally limited. The additional triplet component in the trilayer system generates an additional diamagnetic magnetization not present in the monolayer system. 
In Fig.\ref{fig:comparison}(c) we compare the core size of the trilayer system also with the case of finite $ \alpha_M $ (including the paramagnetic limiting effect). While $ \xi_c $ at $\alpha_M=0 $ is essentially identical to the monolayer case, we see a reduction of the core size with paramagnetic limiting, opposite to the monolayer case [see Fig.\ref{fig:plot1}(c)]. This effect is caused by the increasing importance of orbital depairing for the outer layers which also governs the core size, that is shrinking with higher fields. Fig.\ref{fig:comparison}(d) shows the field dependence of the susceptibility which is more diamagnetic for the trilayer than the monolayer system. 

\begin{figure}
\includegraphics[width=0.49\textwidth]{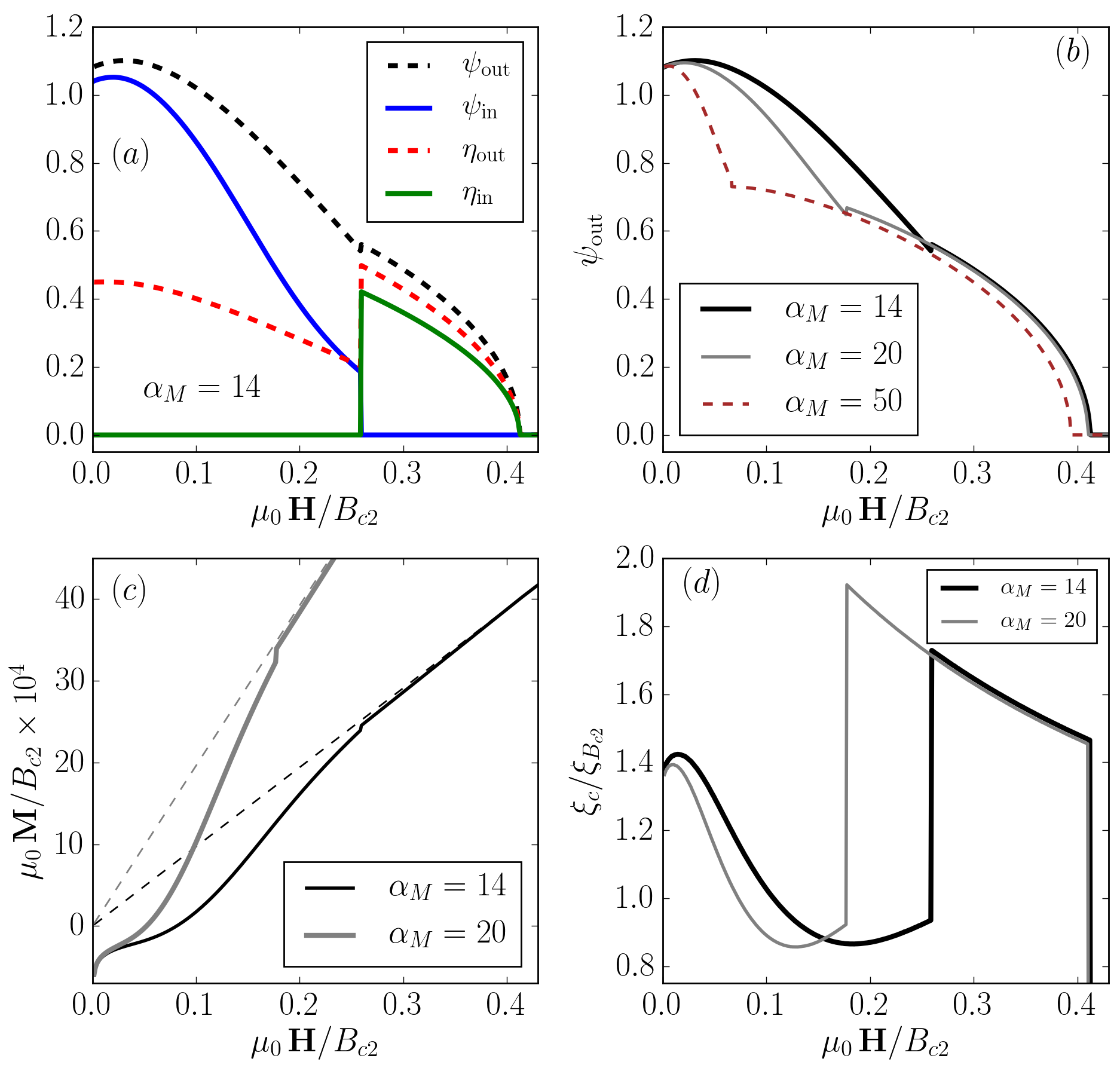}
\caption{\label{fig:orblimited} (a) Magnetic field dependence of the order parameters for a $\alpha_M=14$ system. The upper critical field in this case is about $40\%$ of the $\alpha_M=0$ case, where the upper orbital field is $B_{c2}$. The upper critical field of the PDW phase remains robust around this value, even for much larger $\alpha_M$, as shown in (b). (c) The representative trilayer magnetization for two different Maki parameters with $\kappa_0 = 100$. The $\alpha_M=14$ curve can be compared with the upper left panel. (d) The magnetic field dependence of the representative trilayer vortex core size, normalized with respect to the core size $\xi_{B_{c2}}$. The core suffers a sharp increase through the first order BCS-PDW transition.}
\end{figure}

\subsection{Trilayers subjected to both paramagnetic and orbital limiting}

Discussing now the complete free energy density, we show in Fig. \ref{fig:orblimited} the situation for the trilayer system at $T=0$ within our model, using the Maki parameters $\alpha_M = 14$ and higher, $\kappa_0=100$ and $\gamma = J=1$. We find that the PDW phase appears 
with increasing field for Maki parameters exceeding $10$. At $\mu_0 H/B_{c2}\approx 0.26$ the first-order BCS-PDW phase transition occurs, visible in all order parameters for $\alpha_M = 14$, similar to Fig.\ref{fig:pdw}(b). The paramagnetic effect reduces the upper critical field to well below $ B_{c2} = \kappa_0 $, the bare critical field (see Fig.\ref{fig:orblimited}(a)). 

To illustrate the effect of varying Maki parameters we show the behavior of $ \psi_{\mathrm{out}} $ as a function of magnetic field in Fig. \ref{fig:orblimited}(b) . At low fields the order parameter is suppressed more strongly with increasing $ \alpha_M $. In this way also the critical field for the change to the PDW phase decreases. This sensitivity to the strength of paramagnetic limiting is lost once we are in the PDW phase. Then $ \psi_{\mathrm{out}} $ only depends weakly on $ \alpha_M $ and also the upper critical field around $ 0.4 B_{c2} $ is clearly originating from orbital depairing, as it also varies rather weakly with $ \alpha_M $. This demonstrates well that the PDW phase is a state which is much less vulnerable to paramagnetic limiting.

Fig. \ref{fig:orblimited}(c) displays representative magnetization curves of the trilayer system for two different Maki parameters. Note that a lower $\kappa_0$ value would yield higher magnetization values, because $\chi_n \approx (\alpha_M/\kappa_0)^2/2$. \textcolor{black}{Due to the spin-triplet component in the middle layer in the PDW phase, the magnetization in the PDW state almost coincides with the normal state magnetization (extrapolated dashed lines)}.

\begin{figure}
\includegraphics[width=0.4\textwidth]{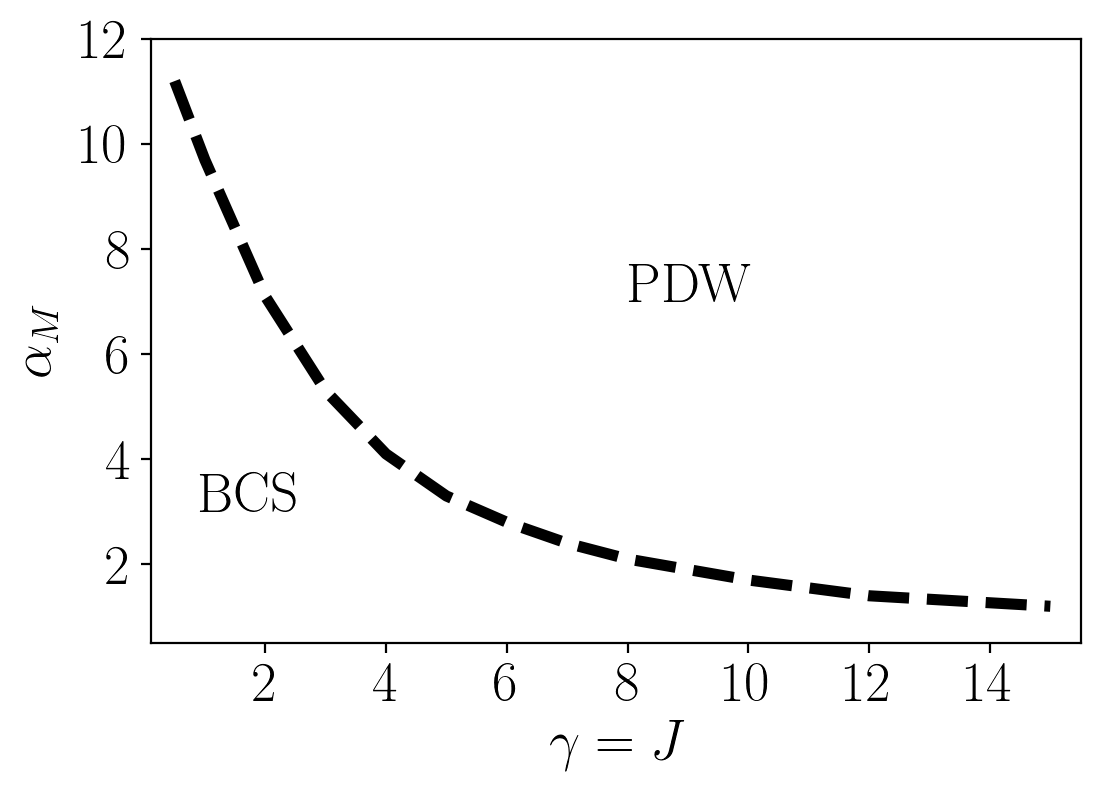}
\caption{\label{fig:ISB} Phase diagram $\alpha_M(\gamma)$ for a superconducting trilayer system. Here we assumed $\gamma=J$, because they should remain comparable. The dashed transition line shows the Maki parameters above which the PDW stabilizes at high fields at zero temperature. The dimensionless energy scale of $\gamma$ might be compared to the superconducting condensation energy, which in dimensionless units is $1/2$. For noncentrosymmetric systems, $\gamma >1/2$ is usually satisfied.}
\end{figure}

The variation of the core size is shown in Fig. \ref{fig:orblimited}(d) for $ \alpha_M = 14 $ and 20. For the BCS phase the effect of paramagnetic limiting has a strong influence on the behavior of $ \xi_c $. At the critical field the core size abruptly increases. That the two curves then basically coincide shows again that 
paramagnetic limiting is essentially irrelevant for the PDW phase. The increase of the core size at the transition to the PDW phase disagrees with the finding in Ref. \onlinecite{Higashi2016} which reports an abrupt decrease based on single-vortex model within a BdG formulation. The discrepancy may have two reasons. First the BdG calculation includes the Kramer-Pesch effect, the shrinkage of the vortex core at low temperatures, which is not captured by the GL treatment\cite{Kramer1974}. Second, for the BdG single-vortex model the core increases with growing magnetic field, lacking the influence of a vortex lattices, in contrast to GL result where the core shrinks as the field approaches $ H_{c2} $. Thus, both approaches have their short-comings and it is difficult to judge, comparing the two treatments, in which way the vortex core would really change at the BCS-PDW transition. However, they have in common that the change is abrupt. 

The threshold for the occurrence of a PDW phase for $ \gamma = 1 $, the case displayed in Fig. \ref{fig:orblimited}, is $ \alpha \approx 10 $. 
With increasing the coupling parameter for the induced parity-mixing in the outer layers, this threshold is lowered. It is the energy gain through the interlayer coupling of the spin-triplet component that helps to stabilize the PDW. The stronger the admixed order parameter $ \eta_i $ the more competitive the PDW phase becomes. We calculate the threshold values for $ \alpha_M $ as a function of $ \gamma $ at $ T=0 $ within our GL model. The result shown in Fig. \ref{fig:ISB} confirms this trend. 

\begin{figure}
\includegraphics[width=0.48\textwidth]{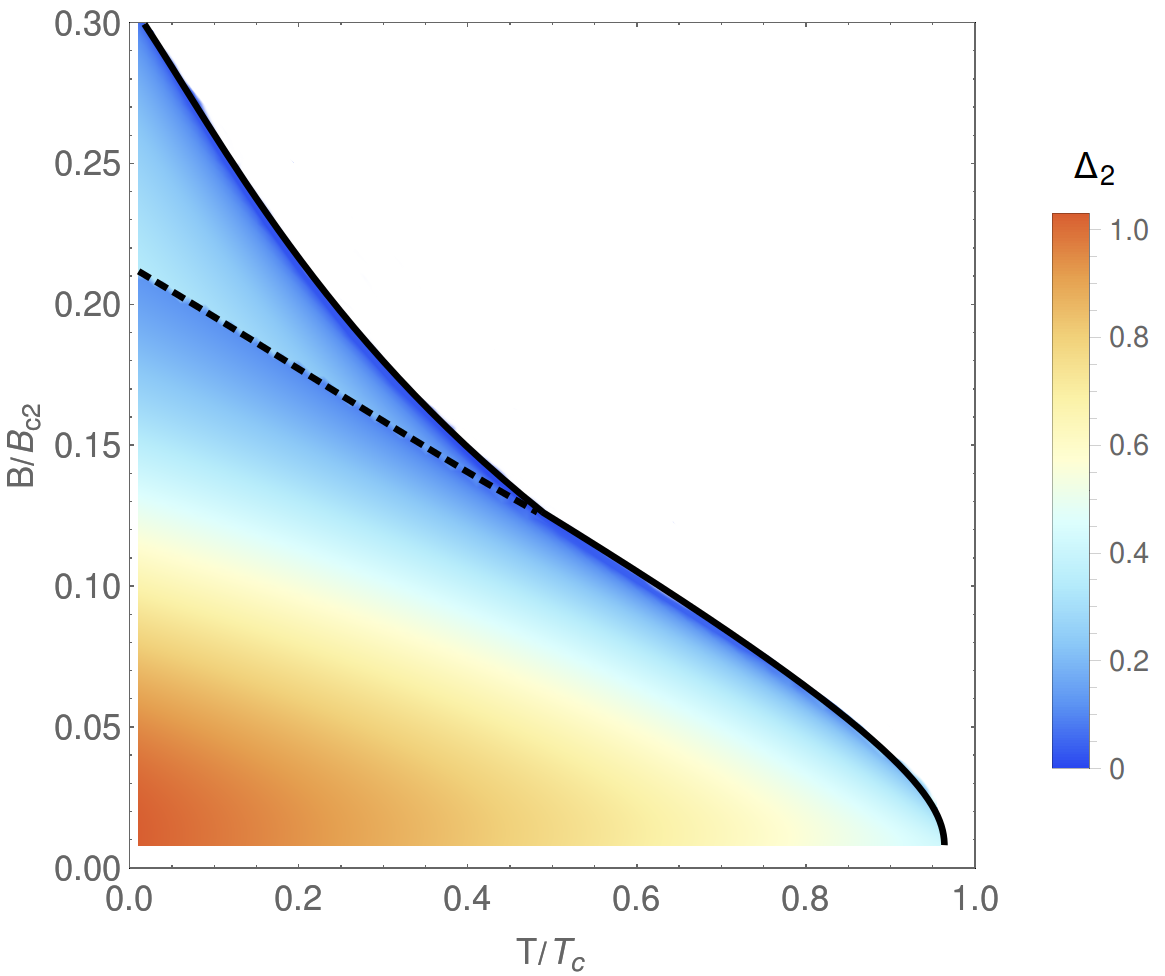}
\caption{\label{fig:finalHT} Typical magnetic induction - temperature phase diagram for the trilayer system with $\alpha_M = 14$, $J=1$ and $\gamma = 3/4$. The color-scale shows the amplitude of the superconducting order parameters in the middle layer, i.e. $\Delta_\mathrm{2} = \sqrt{\psi_\mathrm{in}^2+\eta_\mathrm{in}^2}$. The solid black line indicates a second-order phase transition, whereas the dashed black line shows the first-order BCS-PDW transition. }
\end{figure}

In Fig. \ref{fig:finalHT} we show a typical magnetic field -- temperature phase diagram of the orbitally limited trilayer system. For the sake of presentation, we use a slightly different value for $\gamma$. The magnetic induction axis is normalized with respect to the purely orbital upper critical field $B_{c2}(0)=\Phi_0/(2\pi\xi(0)^2)$. From the $B/B_{c2}$ values we conclude that a rather strong paramagnetic limiting effect is necessary to realize the PDW phase in competition with the orbitally limiting effect. The colors show the amplitude of the superconducting order parameter in the middle layer $\sqrt{\psi_\mathrm{in}^2+\eta_\mathrm{in}^2}$. The solid-black line indicates the second order normal to superconducting phase transition, and the black-dashed line shows the first order BCS to PDW phase transition. 

\section{Final remarks and conclusion}

Interesting properties arise when artificial superlattices including multilayer superconductors are exposed to external magnetic fields. 
An important aspect often ignored is the presence of a local non-centrosymmetricity in these structures that can cause specific forms of spin-orbit coupling giving rise to Cooper pairing with parity mixing. In such a system it was shown that inplane magnetic fields lead to a complex stripe phase of the order parameter resembling a state between the Fulde-Ferrell (FF) and the Larkin-Ovchinnikov (LO) phase\cite{Yoshida2013}. On the other hand, fields perpendicular to the layers can generate a new state at low temperatures and high fields, which may be characterized as a pair density wave (PDW) state. 

In the present study we have addressed this latter situation, extending previous studies using the Bogolyubov-de Gennes formalism without orbital depairing to take into account the penetration of flux lines. This mixed phase is treated within a Ginzburg-Landau model of system of three superconducting layers. For symmetry reasons these three layers behave 
differently. The spin-orbit coupling makes the outer layers more robust against paramagnetic limiting effects in contrast to the center layer. If paramagnetic limiting is important in this material, i.e. the Maki parameter $ \alpha_M $ is sufficiently large, this provides a mechanism to change the configuration of layer-dependent order parameters, from a phase (BCS) where the spin-singlet component of the order parameter is even under reflection at the middle layer to a phase (PDW) where it is odd. This transition upon increasing the magnetic field is of first order.
Increasing the parity-mixing in the outer layers in zero magnetic field lowers the threshold of $ \alpha_M $ for this transition to happen.

The heavy fermion CeCoIn$_5$ compound has an estimated perpendicular magnetic field Maki parameter ranging from $\alpha_M^\perp = 3-5$ \cite{Matsuda2007}. Other heavy fermion superconductors have comparable Maki parameters. Another promising family of superconductors are the iron-based materials. In particular KFe$_2$As$_2$ has an estimated value of $\alpha_M = 15$\cite{Burger2013}, which makes it a strong candidate for the PDW phase in the appropriate superlattice structure. Other iron-based superconductors could also be candidates \cite{Zhang2011}. These materials might due to their layered structure also be suitable to produce multilayer superlattices\cite{Shiogai2016}.

\begin{acknowledgments}
We would like to thank S. Etter, T. Bzdu\v{s}ek and A. Ramires,  for helpful and enlightening discussions.
D. M. is grateful for the financial support and hospitality provided by the Pauli Center of ETH Zurich, and partial financial support from Brazilian agency CNPq,  and the support by the Israel Science Foundation, Grant No. 1287/15.
M.S. is grateful for the financial support by the Swiss National Science Foundation through Division II (No. 163186) and Y.Y. acknowledges financial support through KAKENHI Grants (No. 15H05884, No. 16H00991, No. 15K05164, No. 15H05745).
\end{acknowledgments}

\clearpage

\appendix

\section{Dimensionless units \label{apa}}

In this paper we perform most of the calculations in dimensionless units to emphasize the quantity of free input parameters to the theory, and performed the calculations at "zero temperature", where the PDW properties are more evident. For the plots, we back-scale to dimensionful SI units for better clarity. The use of a Ginzburg-Landau approach in this temperature regime will obviously only lead to qualitative or semi-quantitative results which are, nevertheless, sufficient to understand the essential features and the necessary conditions on materials properties. 

If paramagnetic limiting is neglected, then the free energy can be expressed in terms of a single free GL parameter $\kappa_0$ \cite{saint1969type}. Therefore, one normalizes the following quantities: $\psi(r)\rightarrow f(r) \sqrt{\alpha(T)/b}$, $r\rightarrow \lambda(T)\rho$, and $\mathbf{A}\rightarrow \boldsymbol{\mathcal{A}}\Phi_0/(2\pi\xi(T))$, where $f$, $\rho$ and $\boldsymbol{\mathcal{A}}$ are the dimensionless normalized superconducting order parameter, radial position and vector potential respectively; and
\begin{equation}
\xi^2(T)=\frac{\hbar^2}{2m\alpha(T)}\quad\mbox{and} \quad \lambda^2(T)=\frac{mb}{\mu_0 q^2\alpha(T)}
\label{textbook}
\end{equation}
are the Ginzburg-Landau coherence length and penetration depth (in the absence of Zeeman coupling to the spin) used for normalization, with $\alpha(T)=a(T_c-T)$, where $a$ is a positive phenomenological parameter, and $T_c$ is the even-parity critical temperature. The flux quantum in SI units is $\Phi=2\pi\hbar/q$, where $q=|2e|$ is the positive charge of a Cooper pair. The sign of the charge is conventional. In dimensionless units the flux quantum is $\phi_0=2\pi/\kappa_0$, the vortex unit cell radius $\rho_\mathcal{B}^2=2/(\kappa_0\bar{\mathcal{B}})$, where $\kappa_0=\lambda(T)/\xi(T)$ is the temperature independent Ginzburg-Landau parameter. 

The dimensionless free energy density $\mathcal{F}$ is related to the free energy density $F$ in SI units via
\begin{equation}
F=\frac{\alpha^2(T)}{b}\mathcal{F}.
\end{equation}
As an example we show the conversion of the term describing paramagnetic depairing to dimensionless units. Phenomenologically, this can be most simply included by the term \cite{Bauer2012,P.Mineev2012}
\begin{equation}
F_p=\frac{1}{\pi R_B^2}\int_\circ\mathrm{d}^2r\,\frac{Q}{2}B^2(r)\psi^2(r),
\end{equation}
where the parameter $Q$ regulates the strength of the paramagnetic effect. This parameter is related to the normal state susceptibility $\chi_n$, which is explained in appendix \ref{apb}. Converting $B$ and $\psi$ to dimensionless units we get
\begin{equation}
F_p=\frac{\alpha^2(T)}{b}\overbrace{\frac{1}{\pi \rho_\mathcal{B}^2}\int_\circ\mathrm{d}^2\rho\,\mathcal{Q}(T)\mathcal{B}^2(\rho) f^2(\rho)}^{\mathcal{F}_p},
\label{qeq}
\end{equation}
where $\mathcal{Q}(T)=\mu_0 Q\alpha(T)/b$ is the only temperature dependent term in the integrand of the dimensionless free energy density $\mathcal{F}$. This leads to a temperature dependent Ginzburg-Landau parameter as was shown by Mineev \cite{P.Mineev2012}. If one does finite temperature calculations then one has to reintroduce the $\alpha^2(T)/b$ factor in front of $\mathcal{F}$ and back-scale to dimensionful units.

An important difference with respect to the singlet and triplet components, is that the singlets are strongly paramagnetically limited, whereas the triplets are less or not at all, depending on the field direction.  The paramagnetic effect of singlet superconductors has been investigated in many papers, some of them are cited in Refs. \onlinecite{Houzet2006,Ichioka2007,Michal2010,P.Mineev2012}.

\section{The Maki parameter within Ginzburg-Landau theory \label{apb}}

The parameter $\mathcal{Q}(T)$ in eq.\eqref{qeq} regulating the strength of the paramagnetic coupling is related to the normal state susceptibility $\chi_n$. To see this, we consider the particular case of a superconductor subjected to paramagnetic limiting only, that is, we neglect the gradients of the superconducting order parameter. The free energy can then be written as
\begin{equation}
\mathcal{F}= -f^2+\frac{f^4}{2}+\frac{\mathcal{B}^2}{1+\chi_n}+\mathcal{Q}(T)\mathcal{B}^2 f^2.
\end{equation}
Here $f$ is a spatially constant dimensionless superconducting singlet order parameter.
One can verify that the critical magnetic fields at which superconductivity is destroyed is $\mathcal{B}_p=(1+\chi_n)\mathcal{H}_p=1/\sqrt{\mathcal{Q}(T)}$. Using the relation
\begin{equation}
\mathcal{H} = \frac{1}{2}\frac{\partial \mathcal{F}_{\mathrm{min}}(T,\mathcal{B})}{\partial \mathcal{B}},
\end{equation}
together with the constitutive relations $\mathcal{B}=\mathcal{H}+\mathcal{M}$ and $\chi(T,\mathcal{B})=\mathcal{M}/\mathcal{H}$, we can write
the superconducting magnetic susceptibility as
\begin{equation}
\chi(T,\mathcal{B})=\frac{1+\chi_n}{1+(1+\chi_n)\mathcal{Q}(T)(1-\mathcal{Q}(T)\mathcal{B}^2)}-1.
\end{equation}
Then, we can relate the normal state susceptibility $\chi_n$ to $\mathcal{Q}(0)$ by the condition $\chi(T=0,\mathcal{B}=0)=0$, which yields
\begin{equation}
\mathcal{Q}(0)=\frac{\chi_n}{1+\chi_n}.
\end{equation}
For $\chi_n\ll 1$, $\mathcal{Q}(0)=\chi_n$, which shows that the free energy part involving paramagnetic limiting  can be written in terms of $\chi_n$ as the only free parameter at zero temperature.

A suitable parameter that measures the relative relevance of the paramagnetic effect with respect to orbital limiting is the Maki parameter $\alpha_M$, which takes into account the orbital upper critical field at zero temperature $\mathcal{H}_{c2}(0)=\kappa_0$, and the paramagnetic critical field $\mathcal{H}_p(0)=[(1+\chi_n)\sqrt{\mathcal{Q}(0)}]^{-1}$. Therefore, we estimate the Maki parameter in the Ginzburg-Landau context as
\begin{equation}
\alpha_M=\sqrt{2}\frac{\mathcal{H}_{c2}(0)}{\mathcal{H}_p(0)}=\kappa_0\sqrt{2\chi_n(1+\chi_n)}.
\end{equation}
For the $\chi_n\ll 1$ case, we arrive at a neat formula relating the normal state susceptibility to the Maki parameter and zero temperature Ginzburg-Landau parameter:
\begin{equation}
\chi_n=\frac{1}{2}\left(\frac{\alpha_M}{\kappa_0}\right)^2.
\label{maki}
\end{equation}
Whereas for an orbital limited type-II superconductor there is only one free input parameter $\kappa_0$, an "orbital+paramagnetic" limited superconductor has two input parameters $(\kappa_0,\alpha_M)$.

\section{Expressions for the magnetic field \label{abig}}

For the sake of clarity of the main text, we include the expressions for the magnetic field $\mathcal{H}$ derived from the free energy density for reference here.
From equations \eqref{hfield}, \eqref{intcore}, \eqref{eq:hcapp}, and \eqref{intpar} we obtain the expression for the applied magnetic field as a function of the variational parameters $f_\infty$ and $\xi_c$ at a given  magnetic induction $\bar{\mathcal{B}}$, which also allows us to obtain an expression for the lower critical field $\mathcal{H}_{c1}$. The magnetic field reads:

\begin{widetext}
\begin{equation}
\begin{split}
2\mathcal{H} = &+\frac{1}{2}\kappa_0\xi_c^2 f_\infty^2(1-f_\infty^2)\ln\left(1+\frac{2}{\bar{\mathcal{B}}\kappa_0\xi_c^2}\right)-\frac{f_\infty^2(1-f_\infty^2)}{\bar{\mathcal{B}}+2/(\kappa_0\xi_c^2)}+\frac{\kappa_0\xi_c^2 f_\infty^4}{\left(2+\bar{\mathcal{B}}\kappa_0\xi_c^2\right)^2}+\frac{f_\infty^2\left(2+3\bar{\mathcal{B}}\kappa_0\xi_c^2\right)}{\kappa_0\left(2+\bar{\mathcal{B}}\kappa_0\xi_c^2\right)^3}  \\
& +\frac{1}{1+\chi_n}\frac{f_\infty}{\kappa_0\xi_c}\frac{K_0(f_p\xi_p)I_1(f_p\mathrm{P})+I_0(f_p\xi_p)K_1(f_p\mathrm{P})}{K_1(f_p\xi_p )I_1(f_p\mathrm{P})-I_1(f_p\xi_p)K_1(f_p\mathrm{P})}+\frac{\left(K_1(f_p\xi_p)I_1(f_p\mathrm{P})-K_1(f_p\mathrm{P})I_1(f_p\xi_p))\right)^{-2}}{(1+\chi_n)\bar{\mathcal{B}}\kappa_0^2\xi_c^2\mathrm{P}^2}\\
&+\frac{\bar{\mathcal{B}}f_\infty^2 \mathcal{Q}(T)}{2}\frac{1}{2+\bar{\mathcal{B}}\kappa_0\xi_c^2}\biggr[8+6\bar{\mathcal{B}}\kappa_0\xi_c^2+3\bar{\mathcal{B}}\kappa_0\xi_c^2(2+\bar{\mathcal{B}}\kappa_0\xi_c^2)\ln\left(1-\frac{2}{2+\bar{\mathcal{B}}\kappa_0\xi_c^2}\right)\biggr].\\
\end{split}
\label{hfunction}
\end{equation}
\end{widetext}
Taking the limits $f_\infty\rightarrow 1$ and $\bar{\mathcal{B}}\rightarrow 0$ in equation \eqref{hfunction}, allows us to obtain an expression for the lower critical field, which reads
\begin{equation}
\mathcal{H}_{c1} = \frac{\kappa_0\xi_{c0}^2}{8}+\frac{1}{8\kappa_0}+
\frac{1}{2\kappa_0\xi_{c0}}\frac{K_0(\xi_{c0})}{K_1(\xi_{c0})},
\label{hc1}
\end{equation}
where $\xi_{c0}$ is the variational vortex core parameter that minimizes the free energy for a single vortex, that is $\partial \mathcal{H}_{c1}/\partial \xi_{c0}=0$, from which
\begin{equation}
\frac{\kappa_0\xi_{c0}}{\sqrt{2}}=\sqrt{1-\frac{K^2_0(\xi_{c0})}{K^2_1(\xi_{c0})}}.
\end{equation}
For $\kappa_0\gg 1$ we see that $\kappa_0\xi_{c0}\approx \sqrt{2}$. Equation \eqref{hc1} can also be found in the first paper by Hao-Clem\cite{Hao1991}, where, to our knowledge, the first version of the circular cell method was proposed to our knowledge. 

Similarly, we calculate the applied magnetic field $\mathcal{H}$ for the trilayer system using \eqref{hfield} and the trilayer free energy density \eqref{full3l}.
To shorten the notation for the calculation we define
\begin{equation}
C_{\kappa}(\bar{\mathcal{B}},\xi_c)=1+\frac{\bar{\mathcal{B}}\kappa_0\xi_c^2}{2}\ln\left(1-\frac{2}{2+\bar{\mathcal{B}}\kappa_0\xi_c^2}\right),
\end{equation}
where it is also convenient to calculate
\begin{equation}
\frac{\partial C_{\kappa}(\bar{\mathcal{B}},\xi_c)}{\partial \bar{\mathcal{B}}}=\kappa_0\xi_c^2\left[\frac{2}{2+\bar{\mathcal{B}}\kappa_0\xi_c^2}+\ln\left(1-\frac{2}{2+\bar{\mathcal{B}}\kappa_0\xi_c^2}\right)\right].
\end{equation}
Then the magnetic field $\mathcal{H}$ is:
\begin{widetext}
\begin{equation}
\begin{split}
2\mathcal{H} = & -\left( \frac{\partial C_{\kappa}}{\partial \bar{\mathcal{B}}}\right)\sum_i\psi_i^2+\kappa_0\xi_c^2\left[\frac{5+2\bar{\mathcal{B}}\kappa_0\xi_c^2}{(2+\bar{\mathcal{B}}\kappa_0\xi_c^2)^2}+\ln\left(1-\frac{2}{2+\bar{\mathcal{B}}\kappa_0\xi_c^2}\right)\right]\sum_i\left(\psi_i^4+\eta_i^4\right) \\
& +\frac{2+3\bar{\mathcal{B}}\kappa_0\xi_c^2}{\kappa_0\left(2+\bar{\mathcal{B}}\kappa_0\xi_c^2 \right )^2}\sum_i\left(\psi_i^2+\eta_i^2\right)+\frac{1}{1+\chi_n}\frac{\Delta_i}{\kappa_0\xi_c}\frac{K_0(\Delta_\infty\xi_\Delta)I_1(\Delta_\infty\mathrm{P})+I_0(\Delta_\infty\xi_\Delta)K_1(\Delta_\infty\mathrm{P})}{K_1(\Delta_\infty\xi_\Delta )I_1(\Delta_\infty\mathrm{P})-I_1(\Delta_\infty\xi_\Delta)K_1(\Delta_\infty\mathrm{P})} \\
&+\frac{\left(K_1(\Delta_\infty\xi_\Delta)I_1(\Delta_\infty\mathrm{P})-K_1(\Delta_\infty\mathrm{P})I_1(\Delta_\infty\xi_\Delta))\right)^{-2}}{(1+\chi_n)\bar{\mathcal{B}}\kappa_0^2\xi_c^2\mathrm{P}^2}+\left( \frac{\partial C_{\kappa}}{\partial \bar{\mathcal{B}}}\right)\sum_i \epsilon_i\gamma\psi_i\eta_i\,\mathrm{sgn}_i(0,\pi)\\
& +\left( \frac{\partial C_{\kappa}}{\partial \bar{\mathcal{B}}}\right)\sum_{\langle i,j\rangle}J\left(|\psi_i-\psi_j|^2|+|\eta_i-\eta_j|^2\right) + 2\bar{\mathcal{B}}\sum_i \mathcal{Q}_i(T)\psi_i^2,
\end{split}
\end{equation}
\end{widetext}
where here we did not consider the spatial dependence of the order parameter for the paramagnetic term for simplicity.

Once the free energy density has been minimized for the order parameters at a given $\bar{\mathcal{B}}$, the magnetic field is calculated using the expressions above, from which the magnetization $\mathcal{M}$ and susceptibility $\chi$ are extracted. We do not include the analytical expressions for $\mathcal{M}$ and $\chi$ here. 

\bibliography{bibliography}

\end{document}